\documentclass[conference, comsoc, a4paper]{IEEEtran}
\IEEEoverridecommandlockouts
\usepackage{algorithmic}
\usepackage{amsmath,amssymb,amsfonts}
\usepackage{cite}
\usepackage{comment}
\usepackage{graphicx}
\usepackage{textcomp}
\usepackage{todonotes}
\usepackage{xcolor}
\usepackage{xurl}

\newcommand {\xx}[1]{} 
\graphicspath{{./figures/}}

\def\BibTeX{{\rm B\kern-.05em{\sc i\kern-.025em b}\kern-.08em
    T\kern-.1667em\lower.7ex\hbox{E}\kern-.125emX}}

\begin{document}

\title{Trusted Hart for Mobile RISC-V Security}

\IEEEspecialpapernotice{\rm\small An abbreviated version of this paper is published in Proceedings of TrustCom 2022 \cite{2022Ushakov}. Compared to that publication, the additional material includes sections \ref{sec:TEE Android}, \ref{sec:keystore}, the text related to Figure \ref{fig:architecture1} in section \ref{sec:enclaves}, and sections \ref{sec:secure boot implementation} and~\ref{sec:TH allocation}.}

\author{\IEEEauthorblockN{
V. Ushakov\IEEEauthorrefmark{1}, 
S. Sovio\IEEEauthorrefmark{1}, 
Q. Qi\IEEEauthorrefmark{2}, 
V. Nayani\IEEEauthorrefmark{1}, 
V. Manea\IEEEauthorrefmark{1}, 
P. Ginzboorg\IEEEauthorrefmark{1} 
and J.E. Ekberg\IEEEauthorrefmark{1}}
\IEEEauthorblockA{\IEEEauthorrefmark{1}\textit{Huawei Technologies}, Helsinki, Finland}
\IEEEauthorblockA{\IEEEauthorrefmark{2}\textit{Movial Oy}, Helsinki, Finland}
}


\maketitle
\thispagestyle{plain}
\pagestyle{plain}

\begin{abstract}
The majority of mobile devices today are based on Arm architecture that supports the hosting of trusted applications in Trusted Execution Environment (TEE). RISC-V is a relatively new open-source instruction set architecture that was engineered to fit many uses. In one potential RISC-V usage scenario, mobile devices could be based on RISC-V hardware. 

We consider the implications of porting the mobile security stack on top of a RISC-V system on a chip, identify the gaps in the open-source Keystone framework for building custom TEEs, and propose a security architecture that, among other things, supports the GlobalPlatform TEE API specification for trusted applications. In addition to Keystone enclaves the architecture includes a Trusted Hart -- a normal core that runs a trusted operating system and is dedicated for security functions, like control of the device’s keystore and the management of secure peripherals. 

The proposed security architecture for RISC-V platform is verified experimentally using the HiFive Unleashed RISC‑V development board.
\end{abstract}

\section{Introduction}
\label{sec:introduction}
After almost 20 years of evolution \cite{2019Matala}, the hardware and software security architecture in mobile devices (smart phones and tablets) is now quite stable. We can say that for Android devices across all manufacturers, a ``mobile security stack'' has emerged -- a common security architecture, where different privilege layers in the mobile device are each responsible for well-defined security services. The integrity, trust root and confidential computing needs for these services are provided by a separate, isolated Trusted Execution Environment (TEE), which hosts a trusted operating system  and Trusted Applications (TAs). The TEE and its interfaces are standardized by the GlobalPlatform (GP) organization. In most of the current Android devices the TEE is set up using the Arm TrustZone isolation architecture \cite{TZ2019, ArmTZ64}. TrustZone enables the construction of a split-world design, where the hardware and software is set up with two parallel operating systems, one of them running in full isolation from the other. TEE is part of the Trusted Computing Base (TCB), i.e. the unconditionally trusted part of the mobile device. 

Even though the TrustZone-based TEEs have served the mobile ecosystem well, with really few practical exploits (like, e.g., \cite{QCEE2016}) occurring over the last decade, there is a mounting worry that the attack surface, i.e. the code size of deployed TEEs -- and with it the TCB -- has grown to several megabytes of highly privileged system code, and thereby is too susceptible to attacks from within, e.g., due to memory vulnerabilities \cite{Cerdeira2020}. At the same time, the static, split-world architecture of Arm TrustZone shows its age compared to more modern hardware-based secure enclave architectures (such as Intel SGX \cite{2013Intel}), that offer similar or better levels of memory isolation with a significantly smaller TCB and a configuration model that is dynamic in time. Further, the latest isolation primitives that combine virtualization techniques with memory firewalling and encryption -- AMD-SEV and Arm Confidential Compute Architecture (CCA) \cite{2021Mulligan} -- 
implement hardware isolation guarantees for a secure workloads executing alongside the Operating System (OS) and scheduled by a hypervisor. One of our goals is smooth transition of services implemented for the split-world, Arm-TrustZone model to these newer, more secure platforms, which we call dynamic enclave architectures.

RISC-V is an open source Instruction Set Architecture (ISA) that was engineered to fit many uses \cite{2020Greengard}. The Linux OS has already been ported to RISC-V hardware, and some of the recent high-end RISC-V chips claim to match the capabilities of Arm-based Central Processing Units (CPUs) in mobile devices \cite{Xuantie-910, 2019Prior}. Thus, in one potential RISC-V usage scenario, a mobile device could be based on RISC-V hardware -- in such a setup not only part of the peripheral components, like the RF chips \cite{2019Shilov}, but also the main CPU of the smart phone's System on Chip (SoC), would be based on RISC-V architecture.




In this paper we consider how to port the mobile security stack (based on a split-world hardware design and a TEE) on top of CPU architectures that support dynamic enclaves. We make the following contributions:
\begin{enumerate}
    \item We present a security architecture for multi-core mobile devices with dynamic enclaves that:
    \begin{itemize}
        \item Supports the GP standard interfaces \cite{GP-INT-API, GP-TEE-API}, TEE, and TA mechanisms \cite{GP-TEE-SYS}; and where 
        \item One of the normal cores is dedicated to run a trusted operating system. We name this core Trusted Hart\footnote{The word ``hart''  means ``hardware thread'' in the RISC-V terminology. In most cases, particularly in all currently existing RISC-V implementations, it is an equivalent of a CPU core.}
        (TH) and use it for security functions, like control of the device's keystore and the management of secure peripherals. 
    \end{itemize}
    \item We validate our architecture via prototype implementation and analysis: 
    \begin{itemize}
        \item We implement our design on a RISC‑V development board \cite{HiFive} using the Keystone framework \cite{2020Keystone} for dynamic enclaves and the seL4 microkernel \cite{Klein2009} as the TH OS.
        \item We port the OP-TEE PKCS \#11 service\cite{OPTEE} -- a large, multi-featured GP TA -- to our system for functional validation. 
        \item We measure performance in the prototype setup.
        
    \end{itemize}
\end{enumerate}

\section{Background and Related Work}
\label{sec:backrel}

\subsection{TEE in Android Devices}
\label{sec:TEE Android}

\begin{figure}[ht]
  \centering
  \includegraphics[width=0.35\textwidth, trim=9.5cm 5.5cm 11cm 3.5cm,clip]{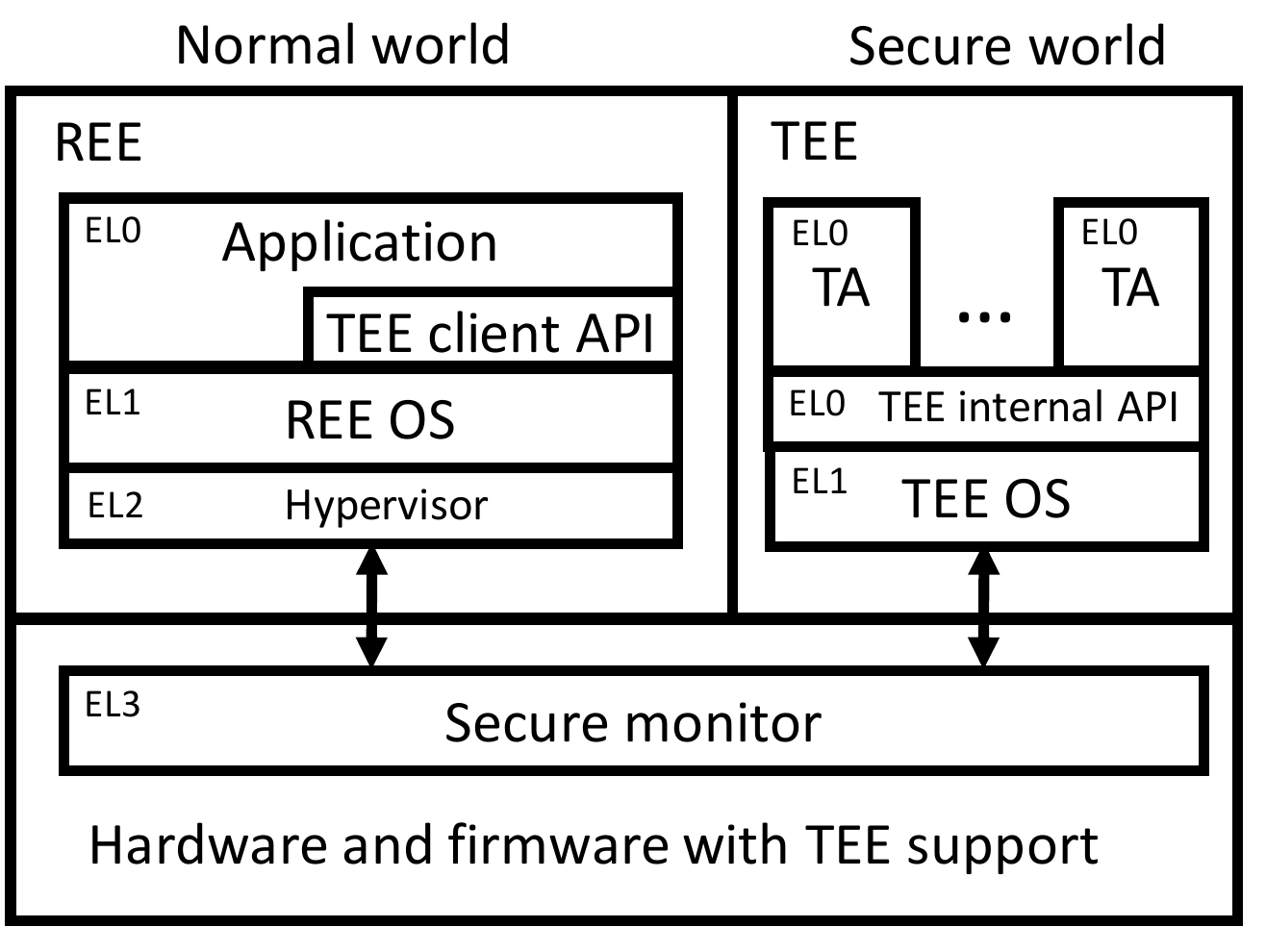}
  \caption{Split-world architecture of Arm TrustZone.}
  \label{fig:stack}
\end{figure}

In most of the current Android devices
the TEE is set up using the Arm TrustZone isolation architecture \cite{ArmTZ64}.
TrustZone enables hardware reuse between Regular Execution Environment (REE) and Trusted Execution Environment (TEE) based on hardware firewalling\footnote{In the Arm architecture one example of a memory firewall is the TrustZone Address Space Controller (TZASC) \url{https://developer.arm.com/documentation/ddi0431/c/introduction/about-the-tzasc}}
 and a set of architectural features. Every bus
access is tagged either as secure or non-secure, providing isolation between the
secure and the normal worlds. CPU state is banked between them and the context
switching is done with the help of the EL3 processor exception (i.e. privilege) level. 
The exception levels EL0, EL1, and EL2 are for user space code, OS kernel, and the
hypervisor, respectively. Figure \ref{fig:stack} depicts this architecture.%

In a mobile phone, the REE OS kernel, e.g. the Linux OS kernel,
together with a few selected system services provides for application isolation,
file storage security, access control and permissions as well as device
integrity.
The TEE
runs a secure TEE OS and some 20--40 Trusted Applications (TAs) that use the
GP  TEE internal API to support secure services such as biometric
authentication, cryptographic keystores for device and service authentication
and attestation, Digital Rights Management (DRM), Trusted User Interface (TUI),
payment applications, etc. The TEE internal API can be compared to a platform
support API, like POSIX; the TAs are implemented against the standardized
features of this API. Already in 2017, more than one billion TEEs were deployed
in mobile devices \cite{trustonic2017}.

The hardware assistance for the TEE protects its OS and services from most common software
attacks, e.g., information leakage. 

The single gateway into the TEE is via the
secure monitor (nowadays called trusted firmware) layer. It is responsible for the
security of TEE invocations from REE. This includes selecting a TA, translating
virtual memory access and copying data between REE and TEE. The secure monitor and the TEE OS are part of the device's TCB.

To ensure the integrity of TEE and TAs, the device's bootloader validates the
manufacturer's signature over the binary of secure world firmware when the
device is powered on. The hash of the ``trust root'' -- the public key of the
manufacturer, which is needed for this signature verification -- is typically
stored in One Time Programmable (OTP) memory (e.g., eFuse).

Device-specific keys that are needed by the TEE are derived from a
Hardware-Unique Key (HUK). The HUK is typically stored in OTP by the
manufacturer and is accessible only from TEE. This ensures HUK's integrity and
protects its confidentiality from REE software. In devices that include
special-purpose hardware (secure co-processor, or cryptographic accelerator) HUK
can be stored in that hardware and is then isolated also from the TEE.

\subsection{Keystore in Mobile Devices}
\label{sec:keystore}

Banking, electronic identity, or DRM application keys in mobile devices need to be protected against malicious access by other applications or by an unauthorized user, who may be present in case the device is lost or stolen. 
Such cryptographic keys have to be protected both when the device is powered on and when the device is powered off.

In Android devices this protection is managed by the Android Keystore system that provides trusted (i.e., persistent and isolated) storage of keys. It also provides an API for developers using the keys. 
As an REE based (software-only) keystore is vulnerable to exposure of keys from attackers' software, the keystore logic as a rule resides in the TEE. A hardware-backed keystore can, e.g., ensure that a key cannot be used without the device.
%
%


\subsection{RISC-V architecture}
\label{sec:RISC-V}

The RISC-V system architecture includes three privilege levels called ``modes''. They may be naturally put in a correspondence to Arm's exception levels. The RISC-V U-mode and S-mode closely  match EL0 and EL1 respectively.
S-mode controls the virtual memory when it is included in the hardware architecture. S-mode is suitable for OS kernel operation.

The RISC-V Foundation has recently added support for hypervisor to the standard.\footnote{ \url{https://github.com/riscv/riscv-isa-manual/releases/download/draft-20211105-c30284b/riscv-privileged.pdf}} However, there are no implementations of this extension yet. 

M-mode is somewhat similar to EL3 in Arm, i.e. the access to many core controlling registers is possible from M-mode only. Particularly the physical memory protection (PMP) registers are only accessible from M-mode. Memory firewalls in RISC-V are based on the PMP.
Besides, M-mode has priority handling of traps and interrupts. 

    
The modern Arm architecture implements the concept of secure mode, where an extra address bus bit indicates the secure or insecure access status. Even though RISC-V does not have such concept, to some extent it can be emulated using PMP registers: the physical address  space can be split into secure and non-secure areas and the current mode of operation can be selected from M-mode by switching accessible PMP areas. 

The advantage of this approach is in its versatility: instead of a simple secure/non-secure
    dichotomy any number of isolated compartments may be created, provided that there are enough PMP registers. For example, in the Keystone framework several separate enclaves 
    may coexist while being isolated from each other as well as from the non-secure world.  

    Every RISC-V implementation includes the M-mode and thus needs a code that
    runs in M-mode. As a bare minimum, this code must initialize the system before
    dropping to a lower privilege mode. Typically, it also implements interrupt 
    delegation and a few simple services for the less privileged code. These 
    services are called Supervisor Binary Interface (SBI).

\subsection{Enclaves}
\label{sec:enclaves}

An enclave is an isolated execution environment put in place to securely process a task required by a host application. The isolation is often set up as a virtualized environment, supported by memory management, memory firewalling and memory encryption done in hardware. The software components that are responsible for setting up the enclave, as well as those that may have access to its memory at run-time form the Trusted Computing Base (TCB).

We define a ``dynamic enclave'' to be an enclave that can be configured, set up and torn down while the device is active. This sets it apart from traditional, more static setups like split-world TEEs and TAs running in them, where the division between secured and non-secured memory is finalized during device boot.

The TCB of dynamic enclaves is as a rule significantly smaller than a full TEE OS kernel -- only enclave setup and I/O are orchestrated centrally for an enclave. A smaller TCB generally implies a reduced attack surface for penetration activities. This approach also adds to service provider flexibility -- enclave code may come with its own support libraries and even run-times, and API lock-in can kept to a minimum. The potential downside of this approach is duplication in memory -- two enclaves using the same run-times or libraries will cause twice the memory to be allocated for this part of the code.

The flexibility of dynamic enclaves comes with the shortcoming that persistent enclave state or secrets cannot be maintained when the enclave is not running and no memory is allocated for it. Still, resource utilization improves significantly as memory need not be divided (and reserved) a priori between secure and insecure workloads independently of actual use. 

A business-oriented distinction between split-world TEEs and dynamic enclaves, is that service provisioning in the former as a rule requires code signatures from the device's manufacturer;  dynamic enclave systems may allow any code to be launched, binding to secrets and authentication using enclave endpoint attestation primitives. 
Emerging enclave systems typically include the generation of a such a signed attestation report by a more privileged part of the system. The receiver of the report can then verify that the enclave code is correct and is running in a secure platform. The attestation report can include the hash of the enclave's code, and other attributes related to the attestation protocol used by the system, e.g., protocol nonces, enclave's identity, code version, and device's boot state.

The Keystone framework \cite{2020Keystone} adds dynamic enclave support to the RISC-V platform. A
Secure Monitor (SM) running in M-mode is able to switch between a ``normal'' workload  and the enclave by 
 reconfiguring the RISC-V memory firewall (PMP) on enclave entry and exit. The process is driven 
 using service calls from either REE or enclave sides. Functions to create, validate and destroy
 an enclave as well as to switch between modes are provided.
A Keystone enclave comprises of kernel and application parts, that run in S-Mode and U-Mode, respectively.  

In this system, an enclave runs in the context of an REE process, i.e. the execution thread of the REE process moves
into the enclave domain when the REE calls the enclave. When inside the enclave, there are 
three ways to return back to the REE:
(1) The control is released back when the requested operation is complete.
(2) The enclave surrenders control to invoke an REE supplied callback to request some service.
(3) A temporary switch from an enclave may be pre-emptively forced by the secure monitor based on timer expiration (to allow REE process scheduling).

A Keystone enclave may request the SM to provide an attestation report signed by the device's private key using the Ed25519 signature algorithm. The report includes (i) hashes of the SM and attestation public key of the SM, both signed by the device's private key; and (ii) hash of the enclave at initialization, and up to 1~KB data block provided by the enclave, both signed by the SM's attestation private key. The receiver of the report can verify the content and the signatures in (i) and (ii), if it has the device's public key, the expected hash of the SM, and the expected hash of the enclave.

\subsection{Related Work}
\label{sec:related work}
Suzaki et al. \cite{Suzaki2020} have proposed a library implementation of GP Internal API for Intel SGX and Keystone enclaves. They divided the GP Internal API into functions that can be implemented as portable library, and functions that depend on the hardware configuration of the device. They do not look further into how to structure the interface between dynamic enclaves and the device's hardware. In their prototype implementation of a subset of the GP Internal API functions, all security functions of the API reside in dynamic enclaves.

Brasser et al.~\cite{2019Brasser} propose to add dynamic enclaves
to the TrustZone-based architecture of Arm devices (see Figure \ref{fig:stack}). They reserve for dynamic enclaves a dedicated ``SANCTUARY Instance'', comprising a CPU core and memory partition, and use the TZASC memory firewall to isolate
dynamic enclaves from each other, the legacy software in the normal world, and the secure world. The SANCTUARY core is launched from the Secure Monitor when needed, and returned to the system after its dynamic enclaves have finished running.
The SANCTUARY runtime dynamically reserves and enforces memory security when an enclave is created.

The CURE hardware enclave architecture of Bahmani et al.~\cite{2021Bahmani} collects several techniques for enclave hardware support into one design, which the authors have prototyped on RISC-V. Enclave memory is configured with dedicated cache ways, thereby mitigating the risk of side-channel attacks. Enclave identifiers are not only included into core memory management, but also added to system bus arbiters for both CPU and DMA controllers. This allows peripherals to be temporally associated with enclaves. 
%
Future will show whether these hardware improvements will be picked up by the mainstream RISC-V specification and chip implementations.

Schneider et al. \cite{2021Schneider} consider the problem of integrating external devices with TEE and propose the notion of ``composite enclave'' -- a set of software or hardware components, called ``unit enclaves''. For example, a composite enclave may include a software unit enclave in the main processor, and a keyboard that is connected to the main processor over the Serial Peripheral Interface (SPI). The latter is a hardware unit enclave. Remote attestation of a composite enclave reports measurements of its software and hardware components. The notion of composite enclave is a possible extension to our architecture with the provision that the attestation of composite enclave is done in TH.

\section{Adversary Model and Architecture Requirements}
\label{sec:requirements}

We assume a powerful adversary that may compromise all software layers including the OS, except for the TEE platform. The goal of the adversary is to leak secret information from a victim TEE workload. This is a common adversary model of TEEs and enclaves \cite{2021Bahmani,2017Subramanyan}.

The requirements for the architecture that we list below include functional requirements F1-F4, and security requirements S1-S5. The latter are  based on the GlobalPlatform TEE System Architecture \cite{GP-TEE-SYS}.

\noindent
{\bf F1.} Fit into RISC-V system architecture. 

\noindent {\bf F2.} Include dynamic enclaves and support for attestation of enclaves.

\noindent {\bf F3.} Support API compatibility of enclave applications with existing GP TEE trusted applications (TAs).

\noindent {\bf F4.} Include a POSIX-compatible REE operating system. This implies Unix-like file system, processes, and access control, and enables source code compatibility of POSIX-compliant applications.

\noindent {\bf F5.} Include hardware-backed (i) isolation of master key, (ii) source of entropy and (iii) monotonic counters.

\noindent {\bf S1.} Isolation of enclave assets from REE applications and vice versa. An enclave is fully isolated from REE applications and other enclaves. (Neither enclave, nor REE application can access each other's assets.)

\noindent {\bf S2.} Trusted storage of assets such as keys and data.

\noindent {\bf S3.} System Integrity: The TCB is launched with secure boot, guaranteeing its integrity and authenticity, both when the system is powered on and when it is powered off. Further, enclave code integrity and authenticity must be guaranteed throughout the lifetime of the system. TCB integrity includes replay protection of its code and data.

\noindent {\bf S4.} Access control is applied between REE applications and enclaves as well as between enclaves and trusted hart applications. Also, enclaves may be granted varying system resource access (e.g., to trusted peripherals, debugging) based on their identity.

\noindent {\bf S5.} Principle of least privilege. An entity must only be able to access the security resources that are necessary to perform its designated task and nothing more. This limits, e.g., the access privileges from an REE application to an enclave. 


Requirements F1-F4 are straightforward.  Note that source code compatibility (F3), i.e. the possibility to compile existing TAs, does not in itself guarantee security. We will discuss F5 in section \ref{sec:hardware backed}. 

Platform security systems existing today tend to satisfy the GlobalPlatform requirements S1-S5 listed above: First, system integrity for secure execution needs to be asserted, e.g., by leveraging secure boot, or verification of integrity by attestation. Second, isolation guarantees need to be upheld for running workloads. Finally, there needs to be a set of hardware-backed services and data that are available for secure workloads. These include secure storage, access to good randomness, roll-back protection, platform-unique and well protected key material for device authentication and encryption.

\section{Architecture}
\label{sec:architecture}

\subsection{Motivation and Overview}

\begin{figure*}[th]
  \centering 
  \includegraphics[width=0.9\linewidth, trim=0 0 0 0 ,clip]{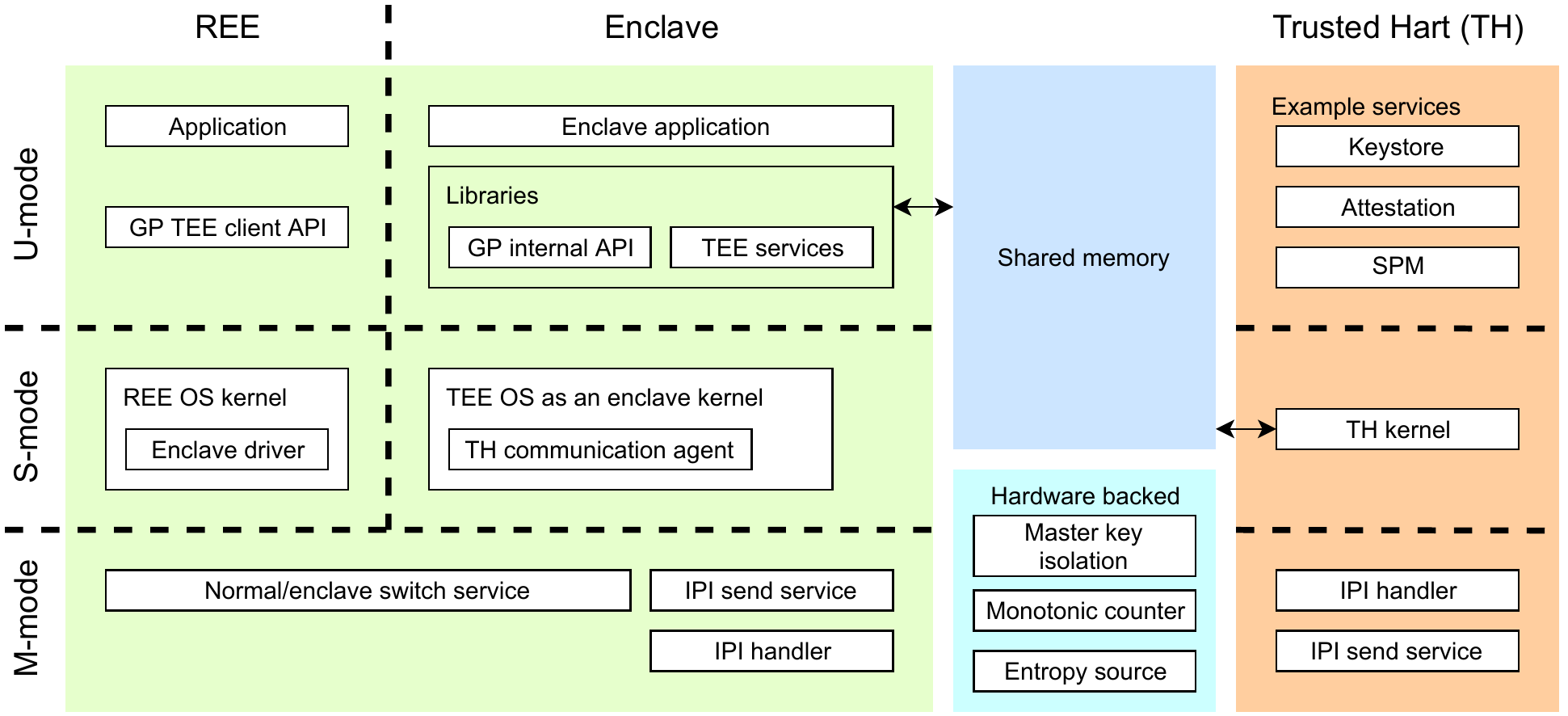}
  \caption{
    \footnotesize Overview of the system architecture. 
    Dashed lines denote the privilege-based memory protection domains.
    Inter-processor interrupts (IPI) are sent using an SBI call and handled on the target hart by redirecting them to S-mode.
    The M-mode code is called Secure Monitor (SM).
  }
  \label{fig:architecture}
\end{figure*}

We want to provide a TEE infrastructure on RISC-V platform that is compatible with GlobalPlatform (GP) \cite{GP-TEE-SYS}. As a starting point we chose
dynamic enclaves defined by the Keystone framework \cite{2020Keystone}. 
But the mapping of GP concepts to Keystone is not trivial. For
example, it would seem natural to map GP TAs to Keystone enclaves.

One similarity between the two is the execution model. The enclave code scheduling
is done by REE with a little help from the SM that prevents the enclave code
from seizing control for too long by processing the secure timer interrupt.
Apart from this the enclave is only executed in response to client requests,
which directly corresponds to the GP request-response execution
pattern.

The differences are more numerous. To name a few, the lifetime of GP TA is managed by the TEE. Also, a GP TA can be persistent, in which case it should be accessible and keep its state during the whole uptime of the device. This is a costly setup in terms of memory and firewalling if implemented with dynamic enclaves.

In addition, the GP TEE also assures the authenticity 
of the secure code, while Keystone merely provides isolation: any code 
may be loaded into an enclave by a REE application and no entity exists to track 
the origins of the secure application. GP client API calls for finding 
a TA in the registry and establishing a session do not easily match the Keystone 
enclave loading model.

Also, a conformant GP TEE provides its TAs with numerous services.
Some of them, such as cryptography primitives, could be implemented as libraries
linked into the enclave code. Others, such as those providing secure persistent
storage or accessing secure peripherals, require a higher degree of protection
than the isolation from the insecure world may provide.

We address the differences between the GP and Keystone concepts by introducing a persistent secure
entity that we call Trusted Hart (TH). It is a RISC-V hart reserved at boot
time and entirely committed to security tasks. 
Since the lifetime of TH spans the whole system uptime, it can
be used as a coordination point between distinct enclaves; it can provide services
relying on persistence; it can be authenticated during the boot and, being such,
given access to sensitive hardware.

On the other hand, keeping the bulk of the secure code in enclaves improves the
system scalability. At its peak the system may devote all its computational
power to the secure calculations.

Having extended the Keystone framework with the TH and augmented it
with a few software and hardware components we obtain the architecture shown in
Figure \ref{fig:architecture}.

\subsection{Dynamic enclaves}
\label{sec:enclaves}

The left half of Figure \ref{fig:architecture} depicts a typical CPU hart that
hosts both REE and enclave environments. Switching between the two is performed
with the help of the Secure Monitor (SM) that reprograms PMP memory areas and swaps
contexts upon an explicit request from the caller via SBI and in accordance with
the enclave metadata it stores. Both environments contain their respective OS
kernels (enclave runtimes), user space applications and libraries; and, whenever
applicable, device drivers facilitating the communication with the SM and/or the TH (see below).

\begin{figure}[ht]
  \centering 
  \includegraphics[width=1.0\linewidth, trim=0 0 0 0,clip]{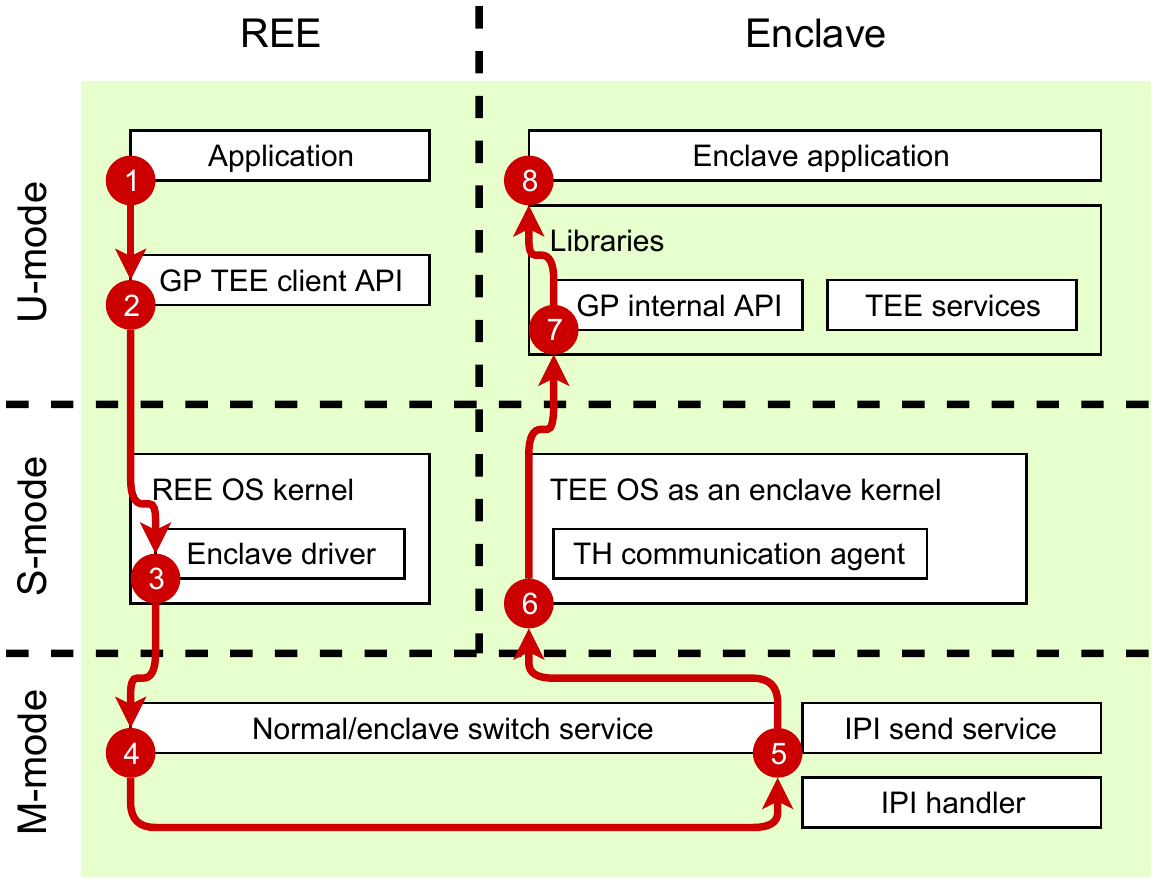}
  \caption{\footnotesize Invocation of enclave from REE application. 
(1) An application requiring a secure service invokes a method.
(2) The TEE Client API support library uses dispatches the call to the driver
    using an OS-specific method.)
(3) The driver stores a request in a buffer designated for the communication
    with the enclave and invokes the Secure Monitor (SM) residing in M-mode.
(4) The SM swaps the PMP setup enabling enclave memory access and
    disabling REE memory, thus effectively switching to the secure enclave.
(5) The SM swaps contexts from Linux kernel to enclave runtime and
    drops into the latter.
(6) From the Enclave runtime's point of view it was a return from an M-mode
    service request marking the end of the previous transaction and the start of
    the next one. After retrieving the accompanying data it drops into the
    enclave application.
(7) The GP Core API library translates the event into a TA invocation.
(8) The TA fulfills the request. The response gets passed
    through the same steps backwards.}
  \label{fig:architecture1}
\end{figure}

The part of the architecture just described matches the Keystone
framework. It satisfies the requirements F1, F2, F4, and S1. On top of Keystone
we add two sets of libraries: one implementing the GP's TEE Client
API \cite{GP-TEE-API} is for the REE side; another one is for linking with the
enclave code -- it implements the TEE Internal Core API \cite{GP-INT-API} and
possibly other TEE service APIs. The libraries satisfy the requirement F3, but to
do so they may require assistance from their counterparts in the TH as
described below.
A typical enclave invocation scenario is illustrated in Figure
\ref{fig:architecture1}. 

\subsection{Trusted Hart}
\label{sec:TH}

The rightmost part of Figure \ref{fig:architecture} represents the Trusted Hart (TH) -- a separate OS installation consisting
of a kernel, a set of device drivers, runtime libraries, and applications that provide
security services. Together with the SM and the bootloader the TH constitutes our TCB and makes up its significant part.  Like REE, the TH makes use of the M-mode firmware via SBI. As its purpose is to serve enclaves, only enclaves are given communication channels to the TH.

\begin{figure}[ht]
  \centering
    \includegraphics[width=1.0\linewidth, trim=0 0 0 0 ,clip]{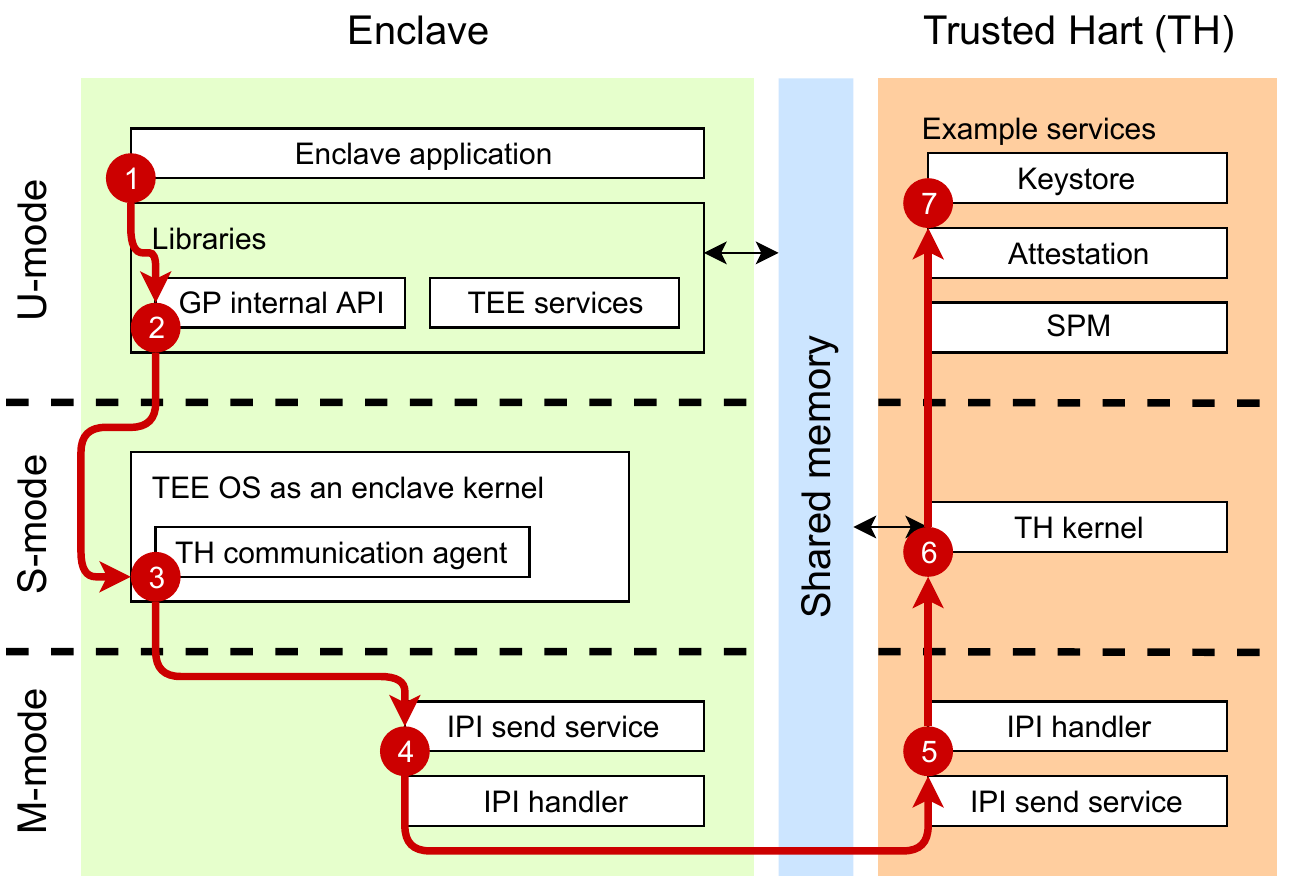}
	\caption{\footnotesize Access to TH from enclave.
(1) A Trusted Application requiring a TH service invokes a method of an
    internal TEE support library.
(2) The TEE support library deposits the associated data into a queue residing
    in the shared memory block dedicated to the TH communication and
    calls the TH support driver.
(3) The driver and the communication agent invoke the M-mode interprocessor
    communication service.
(4) The service uses the hardware means to send an interprocessor interrupt
    targeting the TH.
(5) An interrupt is received by an M-mode interrupt handler in the TH
    and delegated to an S-mode handler.
(6) The TH kernel interrupt handler translates the interrupt into a
    message and dispatches it to the corresponding application.
(7) The application fulfills the request.}
    \label{fig:architecture2}
\end{figure}

TH communicates with enclaves using shared memory buffers and synchronizes them using interprocessor interrupts, as outlined in
Figure \ref{fig:architecture2}. A shared memory buffer for communication with TH is allocated by the SM upon enclave creation from the pool reserved at boot time. The buffers are protected by PMP, which provides confidentiality and integrity of communication. On initial contact the SM passes to the TH the enclave's identity consisting of (i) a hash of the enclave's code measured by SM, and (ii) an external identifier given by the REE client application. Since the SM is trusted, these mechanisms create a secure communication channel between an enclave and TH.   
%

\subsection{Trusted Hart Services}
\label{sec:services}


We will now give three examples of services hosted in TH. Please note that the development of these services should be controlled by the device's vendor, because they are part of the TCB. This makes the set of Trusted Hart services more difficult to extend than the set of enclave applications. For example, updates of the TH and SM are part of the device's firmware update, while installation of an enclave application can be managed in the REE. 

{\bf Keystore. }Trusted Hart is persistent throughout the system run, thus it
can provide a secure key material storage combined with cryptography primitives
to allow non-extractable keys. The keystore is responsible for key generation,
import and usage subject to the access-control policies defined by the
enclave that has requested the key generation or import.

A secure peripheral may be designated for persistent key storage -- it will store the keys 
when the system is down.

{\bf Attestation.} As explained in section \ref{sec:enclaves}, when an enclave requests attestation report in the Keystone framework, the SM creates and signs the enclave's attestation report in raw binary format. The report includes hash of the SM, attestation public key of the SM, hash of the enclave at initialization and optional data provided by the enclave.

When an enclave requests an attestation report in our architecture, the SM sends to the TH the hash of the enclave taken at enclave's initialization and optionally data provided by the enclave. The attestation service in the TH creates and signs the attestation report with attestation key of TH. The attestation key of TH is created at boot time. 

The reasons for making the attestation report in TH, rather than in  SM, are that first, real-world attestation reports come in a variety of content and encoding formats (e.g., JSON, CBOR, ASN.1). Second, privacy-aware report creation may require complex logic. Adding this variety and logic into the SM would increase its complexity and attack surface, making the most privileged part of the system more vulnerable.  

The attestation report includes (i) hashes of the SM and attestation public key of the SM, both signed by the device's private key; (ii) hashes of the TH and attestation public key of the TH, both signed by the SM attestation private key; and (iii) hash of the enclave code at initialization, a data block provided by the enclave, and possibly other fields, signed by the TH's attestation private key. The receiver of the report can verify the content and the signatures in (i) and (ii) and (iii), if it has the device's public key, the expected hash of the SM, the expected hash of the TH, and the expected values in (iii).

{\bf Secure Peripherals' Manager.}
The GP  defines Public, Trusted and Shared Trusted  classes of peripherals: a Public peripheral can be used only from REE; a Trusted peripheral can be used only from TEE; and a Shared Trusted peripheral can be used either from REE or TEE \cite{GP-TEE-SYS}. 

We adapt these in our architecture as follows: a Public peripheral can be used only from REE; a Trusted peripheral can be used either only from an enclave, or only from the TH; and a Shared Trusted peripheral can shared between enclave and TH according to the device manufacturer's policy. The sharing of Trusted peripheral between enclave and REE should be forbidden, because REE is not trusted. A peripheral is assigned to one of these classes at boot time.  

When an enclave requests access  to a Trusted, or to a Shared Trusted peripheral, the SM consults the Secure Peripheral Manager (SPM) in TH, which makes a decision based on the device manufacturer's policy. For example, the SPM may restrict access to the fingerprint reader only to certain enclaves.

Unlike in Arm TrustZone architecture, the Trusted peripheral status cannot be defined on the hardware level in RISC-V. But it can be simulated using PMP zones and M-mode interrupt interception.

\subsection{GP Standard Interfaces}
\label{sec:GP}

Our architecture aims to provide API level compliance with GP
standard interfaces to allow existing client and trusted applications to run in
devices equipped with our architecture after these are re-compiled for the new
target platform.

We support this goal by providing an REE library implementing the GP TEE Client
API \cite{GP-TEE-API} and a library implementing GP TEE Internal Core API
\cite{GP-INT-API}, which is designed to be linked against enclave applications
turning them into GP TAs. 
The latter library uses a set of TH services for
certain GP features, e.g., persistent storage.

\subsection{System Boot and the Root of Trust}
\label{sec:boot}

Secure boot is part of system integrity (S3) mechanisms. In the Keystone framework \cite{2020Keystone} the secure boot starts either from an immutable software, e.g., a Zero Stage Bootloader (ZSBL), or hardware, e.g., a crypto engine. 
When the system starts this entity (1) measures and loads the First Stage Bootloader (FSBL);  which then (2) measures the SM image; (3) generates a fresh attestation key from a secure source of randomness; (4) signs the measurement and the public key with a hardware-visible secret; and (5) stores it to the SM private memory.

In our system, in addition to the above, we also verify and launch the TH during secure boot: the SM (6) measures the TH; (7) generates a fresh attestation key of TH from a secure source of randomness; (8) signs the measurement and the public key with the attestation key of SM; (9) stores it to the TH private memory; (10) launches the TH and the other harts of the system. We describe these steps in more detail in section \ref{sec:secure boot implementation}.

\subsection{Hardware-Backed Mechanisms}
\label{sec:hardware backed}


The requirement F5 mandates certain hardware-backed mechanisms. We leave it open how they are realized in the device hardware platform. 
The rationale for having these mechanisms is the following. 

(i) A mobile device typically includes a hierarchy of keys with a master key at its root. Other device keys may be stored encrypted with a master key. Device security will be broken if that key is compromised. To mitigate this threat, master key should be stored in isolated cryptographic hardware module whose internals cannot be read or modified from outside. Only this module can directly use the master key. 

(ii) Cryptographic keys need to be sufficiently random to prevent an adversary from finding a better search strategy for a key than exhaustive search. Entropy source enables generation of sufficiently random cryptographic quantities in the device by collecting  physical phenomena, which the attacker cannot influence or reproduce, like oscillations in semiconductors, or electromagnetic radiation.

(iii) A monotonic counter is a single machine word in non-volatile memory that can be incremented, but not decremented \cite{2008Schellekens}. It is the basis of local protection in the device against rollback attack, where the attacker attempts to replace, e.g., the device's firmware with an older version.





\section{Implementation}
\label{sec:implementation}

\begin{figure}[t]
    \centering
    \includegraphics[width=\linewidth, trim=0 0 0 0 ,clip]{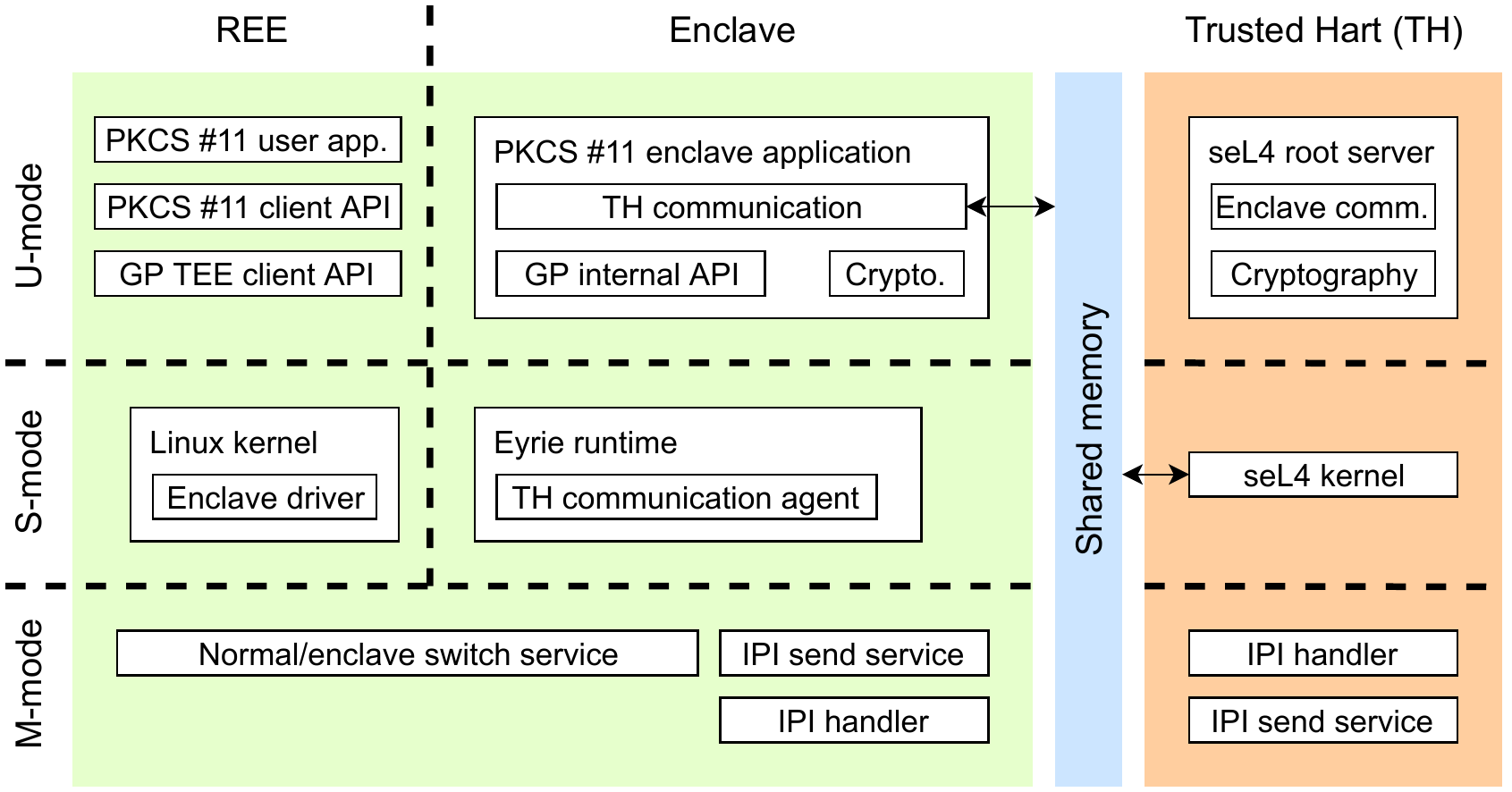}
    \caption{\footnotesize Prototype implementation of the Trusted Hart
                                                                architecture.}
    \label{fig:implementation}
\end{figure}

\subsection{Hardware limitations}
\label{sec:limitations}

The HiFive Unleashed development board that we used as our target hardware is
based on the first RISC-V SoC able to run a full-fledged Linux instance. It has
four fully capable RISC-V harts and a reduced one. The reduced hart lacks a Memory Management Unit (MMU), making it 
unusable for hosting a modern OS. The board also contains 8~GB of DRAM,
Flash and OTP storage, a set of peripherals, and a possibility for expansion. 
The primary focus of its developers was to provide a reference
design rather than a production-ready equipment, so the board lacks a few
security-related features, which are necessary to fully implement our
architectural goals. 

Firstly, ZSBL provided with the HiFive Unleashed FU540-C000 SoC does not verify
the next bootloader stage. Hence, securely establishing the root of trust is not
possible as the FSBL is stored in a dedicated partition on an SDcard and can be
easily replaced.

Secondly, some required hardware-backed components are missing: an entropy source 
and a master key store. 


Finally, PMP registers do provide the memory/device isolation on the hart level,
but cannot protect against non-CPU bus agents. Any actor on the system bus 
bypasses the protection. 


\subsection{Secure Boot Implementation}
\label{sec:secure boot implementation}

The Keystone implementation of secure boot includes derivation of attestation key for the SM and creation of key endorsement token for it. In addition, we derive TH attestation key and key endorsement token for it during secure boot. 

\begin{figure}[t]
  \centering 
  \includegraphics[width=\linewidth, trim=0 0 0 0 ,clip]{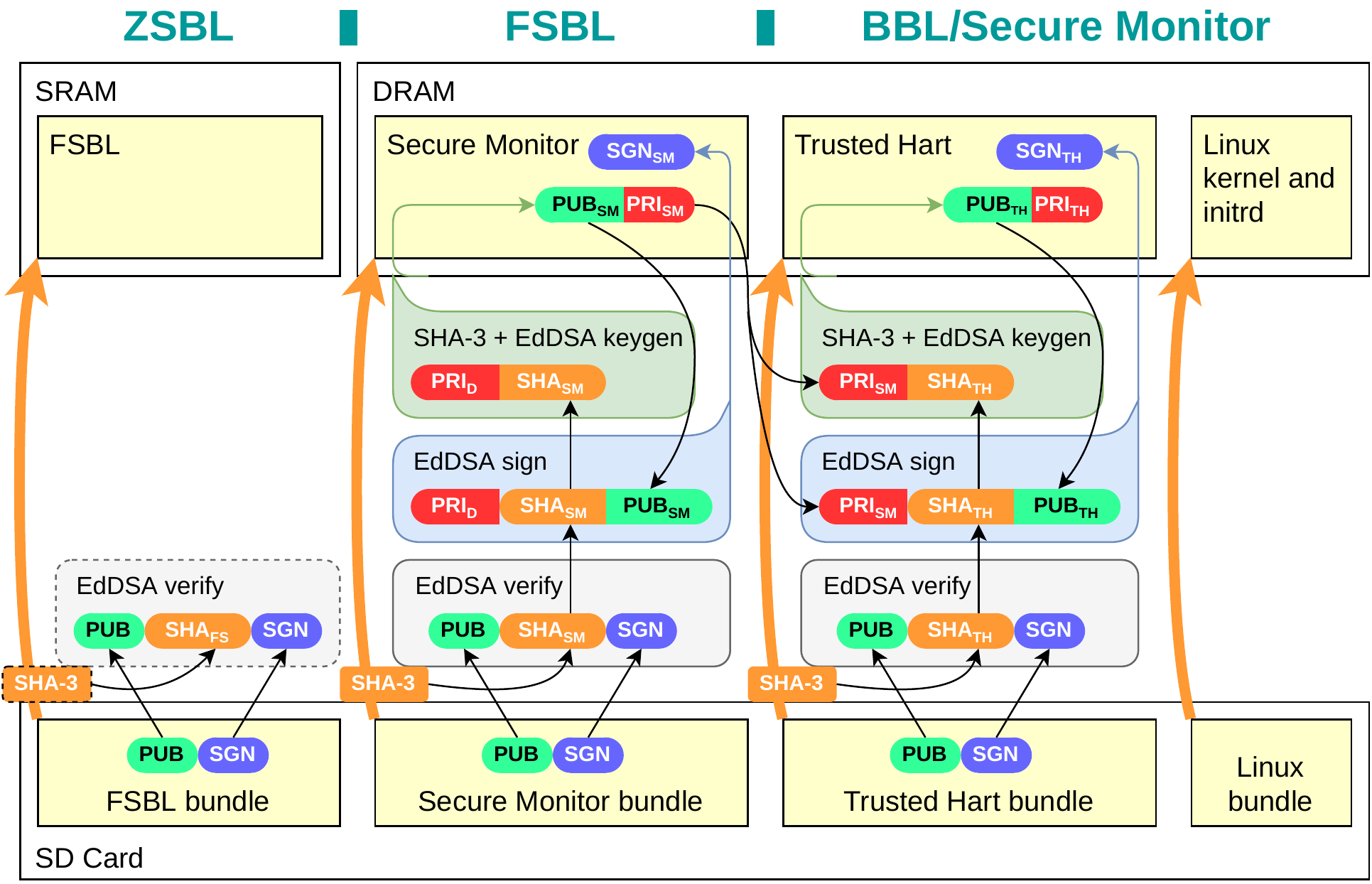}
  \caption{\footnotesize 
The boot process in Keystone implementation has been extended with TH initialization in the BBL/Secure Monitor stage. The Keystone implementation uses EdDSA. 
\newline 
$\bullet$ ZSBL: The ZSBL loads the FSBL from SD card to SRAM, and 
it should verify the signature of the FSBL bundle by using the ``trust root'' of the device. However, the ZSBL in our development board does not verify FSBL. 
\newline
$\bullet$ FSBL: The FSBL first creates the attestation key of SM ($\text{PRIV}_\text{SM}$ in the figure), and then the key endorsement for the public part of that key. This key endorsement consists of ${\text{SHA}_\text{SM} \| \text{PUB}_\text{SM} \| \text{SGN}_\text{{D}}}$, where $\|$ is the concatenation sign, $\text{SHA}_\text{SM}$ is the hash of SM, $\text{PUB}_\text{SM}$ is the public part of the attestation key of the SM, and $\text{SGN}_\text{D}=\text{EdDSA}(\text{SHA}_\text{SM} \| \text{PUB}_\text{SM}, \text{PRIV}_\text{D})$ is the EdDSA signature over $\text{SHA}_\text{SM} \| \text{PUB}_\text{SM}$ with the device private key $\text{PRIV}_\text{D}$, which should be accessible by FSBL. 
\newline
$\bullet$ BBL/Secure Monitor: In our extension of this boot process BBL/Secure Monitor creates the attestation key of TH and the endorsement for that key. The latter consists of ${\text{SHA}_\text{TH} \| \text{PUB}_\text{TH} \| \text{SGN}_\text{SM}}$, where $\text{SHA}_\text{TH}$ is the hash TH,  $\text{PUB}_\text{TH}$ is the public part of the attestation key of the TH, and $\text{SGN}_\text{SM}= \text{EdDSA}(\text{SHA}\_\text{TH} \| \text{PUB}_\text{TH}, \text{PRIV}_\text{SM})$. We currently use the same public key to verify signatures of all bundles.}
  \label{fig:boot}
\end{figure}

We will now outline our implementation of steps (1) to (10) in section \ref{sec:boot}, comprising the secure boot process; the three stages of secure boot are illustrated in Figure \ref{fig:boot}.

{ ZSBL:} (1) The Zero Stage Boot Loader (ZSBL), which is hardwired into on-chip mask ROM, loads the
        First Stage Boot Loader (FSBL) from the Flash memory and it should verify its integrity
        and authenticity.  A real-life ZSBL can verify the FSBL signature using a public key provided together with FSBL code. The public key itself, i.e. the ``trust root'' of the device, is typically authenticated by comparing its hash with the reference value stored in the on-chip OTP memory by the manufacturer, as mentioned  in section \ref{sec:TEE Android}. 

{ FSBL:} (2) The FSBL loads and verifies the SM
        code. (3) It also derives and (4) endorses
        the attestation key of the SM, using the method shown in Figure
        \ref{fig:boot}. 
        (5) The FSBL then passes the attestation key of the SM together with its endorsement data to the next
        bootloader stage contained in the SM.

{BBL/Secure Monitor:} (6) The SM code -- originally based on the Berkeley Boot Loader
        (BBL) -- loads and validates the TH bundle containing seL4 kernel
        and its root server. (7) The attestation key of TH is derived and
        (8) endorsed using the method shown in Figure
        \ref{fig:boot}.  (9) The SM then passes the attestation key of the TH together with its endorsement data to the TH. 
%
All bootloader stages are entered by all harts simultaneously, however all the
job is done by hart 0 alone; other harts are paused until the control is
transferred to the next stage. (10) Finally, the TH is started on hart 1,
harts 2--4 host Linux. Hart 0 is halted after booting.

Please note that the REE bundle consisting of Linux kernel and the initial RAM disk is not validated as it is considered inherently insecure.

TH gets the TH private attestation key endorsed by SM and stores it as a
non-extractable PKCS \#11 private key object.

As we mentioned above, the verified boot process cannot be implemented securely in our development board 
as ZSBL is non-replaceable and it does not verify the FSBL. 

As in the Keystone implementation, the chain of trusted device keys is simulated from a hard-coded device key, instead of being generated
securely. The SM key generation is done properly, but it depends on a secure device key provided by
FSBL, which is simulated.


\subsection{Keystone Modifications}
\label{sec:keystone modifications}

Our implementation is based on a modified Keystone project,
which includes an M-mode layer consisting of a bootloader and a secure monitor,
a Linux kernel driver module, a Linux user space support library, an enclave
runtime and a set of enclave support libraries.%
\footnote{In the beginning of 2021 the Keystone project M-mode layer was
          refactored to utilize the OpenSBI project as its base 
          (\url{https://github.com/riscv-software-src/opensbi}). Currently,
          our code is still forked from the old Keystone based on Berkeley Boot
          Loader.} 
Components of Keystone have been modified as follows:

The bootloader has been modified to allow for the loading of a second operating
system, which acts as a TH runtime. It is started on a dedicated hart,
while Linux (acting as REE in our case) occupies the others. REE is given a
patched device tree where the reserved hart and memory are marked.

The SM has been modified to use a second Universal Asynchronous Receiver Transmitter (UART) for TH and
SM logs; pass and handle inter-hart interrupts to synchronise the
communication between the TH and REE/enclaves (the communication
itself is done directly using the shared memory buffer allocated at the boot
time); and to include the TH measurement data into the enclave
attestation hash. Also a few shortcomings of Keystone implementation have been
corrected, e.g., register loss in the timer based enclave scheduler.
We have also written a new simplified Keystone kernel driver with a renewed physical
memory allocation scheme.
Furthermore, the Linux user space Keystone library has been refactored, to
improve memory management and to e.g., support running multiple enclaves
simultaneously.


The ability to communicate with TH has been added to the enclave
runtime, also enclave interrupt handling has been externalized to the SM. Inter-processor
interrupts from TH are not currently implemented, instead enclaves poll for the
TH operation status on each entry (effectively on every scheduling time slice).

The Keystone implementation reserves one memory zone per enclave; in most cases one zone requires two PMP registers. In total, the Keystone implementation uses two PMP registers per enclave.  We allocate also another memory zone per enclave for communication between that enclave and TH. In total, we use four PMP registers per enclave.  Since our development board has only eight PMP registers,\footnote{Different versions of the RISC-V architecture define different limitations on the amount of PMP registers. The architecture our development board conforms to limits their number to 16, the most recent version allows up to 64.}  and two registers are reserved for system purposes, we can have at most one enclave per normal hart. Note that the limit affects all existing enclaves, not only those that run simultaneously. We discuss how to solve this issue in section \ref{sec:PMP limitations}.


\subsection{GP API Implementation}
\label{sec:gp implementation}

On the top of the modified Keystone code we have added two libraries implementing
GP APIs. One is a Linux shared library providing GP TEE Client API
\cite{GP-TEE-API} for TEE client applications. Another is a static library
intended to be linked against the enclave application code. It implements GP TEE
Internal Core API \cite{GP-INT-API} and allows porting GP TAs. It interfaces to
the TH too, because part of GP logic is handled there.


\subsection{SeL4-based Trusted Hart}
\label{sec:impl:sel4}

SeL4 is a microkernel supported by the seL4 Foundation \cite{sel420200311}. It
belongs to a family of microkernels called L4.
%
Akin to all microkernels seL4 provides exceptional isolation between OS components
at the cost of performance. In addition, seL4 is designed to be formally verified, which
in fact has been done for Arm and RISC-V architectures \cite{Klein2009}. We have
chosen it as the TH runtime environment.

The seL4 kernel has been extended to handle inter-processor interrupts coming from harts not controlled by seL4.
Also seL4 bootstrap process has been simplified: the original module for parsing
and loading of Executable and Linkable Format (ELF) files for the kernel and the root server process has been
removed. ELF parsing is done at compile time, and the loading is handled by
the Keystone bootloader. Formal verification of the modified kernel is left for the future.


\subsection{PKCS~\#11 Implementation}
\label{sec:impl:pkcs11}

As an example service we have ported an implementation of the PKCS~\#11 standard
from the OP-TEE project \cite{OPTEE_os, OPTEE_client}. This involved adapting
the Linux side library providing the PKCS~\#11 API to the application, as well
as the GP TA implementing a virtual PKCS~\#11
token. With this API, applications can address cryptographic devices as tokens and do cryptographic functions implemented by these tokens. 
We also wrote a simple test program, used e.g., for performance testing.

In order to illustrate the features of our architecture
we have added the Elliptic Curve Diffie-Hellman (ECDH) key exchange mechanism in TH to OP-TEE
PKCS~\#11. A private EC key marked ``sensitive'' (implying its
unextractability) never leaves the TH memory; the EC operations with that key 
are done within the TH-hosted code. Keys that are not marked ``sensitive'' are handled by the enclave alone and can be extracted.

All in all, two example services have been partially implemented in TH. 
The keys stored in TH may persist beyond the enclave lifetime, but not
through the system downtime, constituting partial implementation of the Keystore service.
The system generates the TH attestation key during boot time, and this key is available for application usage as PKCS-\#11 object, which partially 
implements the Attestation service.

To save time we have implemented the TH functionality, including the PKCS~\#11 
support, in the seL4 root server application.\footnote{
    Keeping all the required functionality within a single program does not
    fully reveal the seL4 potential.}
    


\section{Rationale and Evaluation}
\label{sec:rationale}


\subsection{Alternatives to Trusted Hart}
\label{sec:altertnatives to TH}

Different (than in TH) options for placing persistent TAs include the following:

$\bullet$ Make the persistent TAs run as a long-lived dynamic enclaves. But this is a costly setup in terms of memory, due to duplication of code in the runtimes of long-lived enclaves, and firewalling -- after the allocation of PMP registers for the long-lived enclave, there are fewer PMP registers for other purposes.

$\bullet$ Make the persistent TAs part of Secure Monitor (SM). This increases the size of  (and thus the potential for vulnerabilities in) the SM, the most privileged part of the device's runtime system. Besides, RISC-V M-Mode does not have an MMU, which increases the chance of memory management errors by developers.

	
$\bullet$ Place the persistent TAs in a separate security processor, like Apple's Secure Enclave Processor (SEP), that is physically isolated from the other cores and can be accessed only via a dedicated hardware inter-process communication mechanism.  But making a fully isolated {\em and} powerful core is expensive. In order to save costs, the separate security processor is typically (i) simpler that the normal core, and (ii) it shares resources, like memory, with the rest of the system.  Both (i) and (ii) may lead to vulnerabilities.\footnote{
Two examples of known vulnerabilities caused by (i) are the buffer overflow and lack of memory protection in Google Titan \cite{GoogleTitanCVE,2021Rossi,2022Melotti}, and an optimisation in Infineon's TPM that caused leakage of its RSA key \cite{2017-ccs-nemec}. 
An example of vulnerability caused by (ii)) is the flaw in the booting process of current Apple's SEPs, which allows attackers to boot an unsigned code in the SEP and access critical keys stored in the SEP \cite{AppleSPUFault}.} 

A security co-processor provides good protection from side channel attacks, and we are not arguing against having a security co-processor in the system, for, e.g., storing device keys. In our view a security co-processor should handle relatively simple tasks, like those of a smart card, which typically handles signatures, and the storing of private keys and counters.  Handling these tasks in a co-processor would also help in supporting multiple hardware platforms. A co-processor will be one of the secure peripherals in our architecture.

\subsection{Static vs. Dynamic Allocation of Trusted Hart} 
\label{sec:TH allocation}

In our implementation the role of TH is allocated statically: at boot time one of the cores of the device becomes TH, and it remains in that role as long as the device has power. A disadvantage of static allocation is that in a device having, say eight cores, dedicating one core for security functions means that one eighth (12.5\%) of the device's computing power is not available for other purposes. 

The TH could be allocated dynamically for more efficient use of device's computing power: if TH is not needed, the system saves the state of TH to secure memory; later, when the TH is needed, it loads that state back into one of the available cores. But the occasional increase of TH response when it is loaded from memory will increase the variation in response time of security services that use TH. Work on this topic is left for the future.

\subsection{Enclave count and PMP limitations}
\label{sec:PMP limitations}

As we have already seen in section \ref{sec:keystone modifications}, the amount of
simultaneously existing enclaves is severely limited by the amount of PMP
registers available in the system. In our implementation we are
unable to load more than one. The problem is that every enclave
defines a few protected memory zones and every protected zone typically needs two PMP
registers. In certain cases one memory zone may be defined by just one register or a
set on $m$ consecutive memory zones may be defined by $m + 1$ registers, but this
does not help much as the number or registers available is small.

The authors of Keystone framework mention two solutions to this problem: (i) adjacent allocation by the OS of memory regions for the enclaves; (ii) support of reallocation by the SM of memory regions, which will allow defragmentation \cite{2020Keystone}. Both solutions are based on the idea that several enclaves may share a protected memory zone if their memory regions are laid out consecutively. 
The downside of solution (i) is that it disallows memory reclamation until all latter enclaves are destroyed. Solution (ii) requires complex interaction between SM, Linux kernel and enclave runtimes.

We suggest to solve the problem   
by keeping in memory a map of protected memory zones, and using PMP registers as
a cache akin to Translation Lookaside Buffer (TLB) tables in virtual memory systems. The SM would need to process the
memory protection faults and update PMP registers when
necessary.  We believe the performance cost of this solution to be adequate, considering the success of the TLB model.

A more serious PMP limitation comes from the fact that PMP areas are defined per
hart and control this hart alone, disregarding the external masters that may
be present on the bus. For instance, HiFive Unleashed board we use includes a
DMA controller and an Ethernet device. Both can access the system memory
bypassing the PMP settings. 

Closing this breach  by virtualizing the bus-mastering devices would impair performance and
significantly increase the complexity of the M-mode layer. 
Besides, the bus may be exported via an external
connector, which renders this approach entirely infeasible.

An effective solution is to put all external bus masters behind secure
bus bridges providing the isolation. Other CPU architectures support Input-Output
Memory Management Units (IOMMU) that can be used for this purpose.  In the RISC-V
community there have been some discussion about extending the PMP architecture
to external bus masters -- they call it IOPMP\cite{2021Ku}. 

\subsection{Performance Measurement}
\label{sec:measurements}

%
%
%
%
%
%
%
%
%











We wrote an REE test application that uses our PKCS \#11 TA. We report timing on two services -- EC key generation and ECDH key exchange. In the ECDH case, the REE simulates two communicating parties, such that each generates an EC key and then performs session key generation with the other party, followed by message encryption and decryption.
The measured execution times generating an EC key pair and running the ECDH exchange are summarized in Table \ref{tab:HiFive}, which shows timed averages over 1000 sample measurements on HiFive Unleashed board.

\begin{table}[ht]
\caption{Measurement results on HiFive Unleashed board.}
\footnotesize
\centering
\begin{tabular}{|c|c|c|}
\hline
                 & EC key pair generation (ms) & ECDH exchange (ms) \\
\hline
Enclave alone   &  12                  &  48         \\
Enclave and TH  & 100                  & 100         \\
of which TH     &  12                  &  23         \\
\hline
\end{tabular}
\label{tab:HiFive}
\end{table}

The measurements were executed in two modes. In case of ``enclave alone'' all the processing is done within an enclave. The ``enclave and TH'' row shows the processing time when the enclave does only the symmetric cryptography, while the asymmetric key pair generation and ECDH exchange is done within TH. The pure TH computational time is shown in the last row.

Observe that in the ``enclave and TH'' row, the measured times of EC key generation and of ECDH operation are both 100 ms. The reason is that in our implementation enclaves poll for the TH operation status on each entry; effectively on every scheduling time slice, which is set to 100 ms.

We take special care to ensure that the computational part of EC primitives does not depend on the environment: We use the same cryptographic library with the same compiler options in both cases. The difference between ECDH exchange times between the ``enclave alone'' and ``of which TH'' rows is due to the differences in the key storage formats: the enclave code stays conformant to the GP specification, while its TH counterpart does not have to.

The measurements on HiFive Unleashed board are the closest metrics we have for a real RISC-V deployment of our architecture. 
As we are internally leveraging QEMU on X86 as a platform for TH development, we also provide timed averages over 1000 sample measurements for this setup in Table \ref{tab:QEMU}. The better performance in this case is mainly an indication of a more powerful CPU, and security shortcuts in the QEMU setup.

\begin{table}[ht]
\caption{Measurement results in QEMU on AMD Ryzen 5950X.}
\footnotesize
\centering
\begin{tabular}{|c|c|c|}
\hline
                 & EC key pair generation (ms) & ECDH exchange (ms) \\
\hline
Enclave alone   &  5                   & 23          \\
Enclave and TH  & 13                   & 20          \\
of which TH     &  9                   & 14          \\
\hline
\end{tabular}
\label{tab:QEMU}
\end{table}

\subsection{Informal Security Analysis}
\label{sec:analysis}

Our architecture builds on top of the Keystone framework, and so we inherit the threats analyzed and the mitigations proposed by that work  \cite{2020Keystone}. Like Keystone framework's authors, we assume that the hardware works as specified, and that secure boot is configured to guarantee the integrity of the TCB. 

The TCB in our architecture includes the bootloader, SM and TH, and we assume that these components function correctly. During secure boot, our architecture moves derivations of device secrets into designated TCB memory locations and shields off further access to them using memory protection. 

We do not mitigate architectural attacks that may affect the integrity of the TCB, e.g., Rowhammer-style attacks, or side-channel attacks leveraging speculative execution. We also do not protect against physical attacks on the device, or physical side-channels like analysis of electro-magnetic field. 

Like most of the earlier work on TEE, this paper does not fully address peripherals / DMA as an attack vector. In our opinion, this can be efficiently mitigated only with IOMMU support in hardware -- and we hope future RISC-V designs will follow ARM in this respect. We also point to related and complementary work \cite{2019Brasser,2021Bahmani} to address peripheral access from enclaves.





Our design fulfills the security requirements in section \ref{sec:requirements} as follows.

{\bf S1.} The Keystone framework isolates enclave assets from REE applications by running all enclaves in hardware-firewalled memory.  

{\bf S2.} We achieve asset protection with the help of the TH, which isolates the management of device secrets and device-specific trust roots from the other enclaves, and also orchestrates the identification of the enclaves. 
The TH manages some assets by itself, and provides enclave access to these only via an authenticated channel.  
Further, it will also manage (distribute) assets on behalf of enclaves using run-time attestation -- i.e. when an enclave starts, its identity is confirmed with help from the SM. 
The TH can safe-keep some assets\footnote{The TH can wrap enclave assets using a dedicated key, which is accessible only by TH.} for an identified enclave, and release these assets to the enclave when that enclave starts running. 

{\bf S3.} System integrity is protected in a traditional way: The TCB of the system -- that includes the SM and the TH -- is securely booted. The REE is not integrity-protected in our design. Each enclave is measured at start-up, and this measurement is used for attestation and enclave access control (to assets) only. 

{\bf S4.}  Access control to enclaves from REE can be implemented in the REE based on enclave identification, but the resulting separation is only as strong as the REE system (which we assume to be unsafe). 
Enclaves are isolated from each other, and they cannot communicate directly. Establishing a cryptographic secure channel between enclaves with the help of TH, so that they can interact via the host is possible, but left for future work. TH is access controlled from enclaves, and the enclave identity will designate, e.g., access to secure storage or derived secrets per enclave.

{\bf S5.}  Enclave access to TH is authenticated based on hash of enclave's code measured by SM. Host application authenticity is handled based on sessions, i.e. the application launching the enclave gets the communication session (file handle) for itself. However, if file handles are shared between processes, then this limitation is easily circumvented. The principle of least privilege therefore hinges on correct enclave and host application implementation. 

We minimize any additional requirements on the SM and concentrate all extra functionality to the TH. This is intentional, we externalise most platform security services to TH. 
TH is validated during secure boot, it is part of TCB and it handles sensitive device assets. In this way the TH services remain isolated from SM -- the most privileged part of the system. This setup allows us to consider also other technical approaches for the TH, such as using embedded hardware.

\section{Final Remarks}
\label{sec:final}



The GlobalPlatform (GP) APIs are widely used in mobile industry; they need to be included also in future mobile devices for supporting legacy Trusted Applications (TAs). 
We have designed and evaluated a security architecture for RISC-V devices that offers a migration path for legacy TAs to this new hardware platform. The architecture includes GP APIs, dynamic enclaves and a dedicated, Trusted Hart (TH) that handles security-critical resources such as device keys, attestation reports and stateful storage. 

As part of evaluation, we wrote a prototype implementation of our architecture on RISC-V development board, and ported an implementation of the PKCS \#11 standard from the OP-TEE project \cite{OPTEE} to this environment.
 
Two limitations of the PMP mechanism in RISC-V  that surfaced during our implementation work are: (i) scarcity of PMP registers, which limits the amount of simultaneously existing dynamic enclaves,  and (ii) the fact that PMP areas, which are defined per hart, do not protect against attacks from peripherals.  We suggest to solve (i) by building a memory area caching system (akin to TLB tables for MMU-enabled system), which allows the count of concurrent enclaves to be limited by available memory only.  There is an ongoing discussion in RISC-V community about IOPMP feature \cite{2021Ku}, which is intended to solve (ii). Despite these limitations in the current RISC-V architecture, we conclude that the RISC-V security mechanisms already can be used to build a TEE for mobile devices.




\bibliographystyle{IEEEtran}
\bibliography{biblio}

\end{document}